\documentclass[12pt]{article}
\usepackage{latexsym,amssymb}
\date{}
\begin{document}
\def\be{\begin{equation}}
\def\ee{\end{equation}}
\def\R{\Bbb R}
\def\C{\Bbb C}
\def\Z{\Bbb Z}
\def\d{\partial}

\def\bbbp{{\rm I\!P}}   
\def\bbbn{{\bf N}}      
\def\bbbO{{\cal O}}     
\def\bbbF{{\cal F}}     
\def\bbbM{{\cal M}}     
\def\bbbL{{\cal L}}     

\renewcommand{\theequation}{\thesection{.\arabic{equation}}}
\newtheorem{definition}{Definition}
\newtheorem{theorem}{Theorem}[section]
\newtheorem{proposition}[theorem]{Proposition}
\newtheorem{lemma}[theorem]{Lemma}
\newtheorem{corollary}[theorem]{Corollary} 
\centerline{\Large \bf On the explicit solutions} \medskip
\centerline{\Large \bf of the elliptic Calogero system} \vskip 1cm
\centerline{\large {\sc L. Gavrilov}}
\centerline{\footnotesize\it Laboratoire Emile Picard , CNRS UMR 5580,
Universit\'e Paul Sabatier}
\baselineskip=12pt
\centerline{\footnotesize\it 118, route de Narbonne, 31062 Toulouse
Cedex,
France}
\centerline{\footnotesize E-mail: gavrilov@picard.ups-tlse.fr} \medskip
\centerline{\large {\sc A.M. Perelomov}
\footnote{On leave of absence from Institute for Theoretical and
Experimental
Physics, 117259 Moscow, Russia. Current e-mail address:
perelomo@posta.unizar.es} }

\centerline{\footnotesize\it Max-Planck-Institut f\"ur Mathematik,
Vivatsgasse 7, 53111 Bonn, Germany}
\medskip

\medskip
\centerline{\sc Abstract}

{\small\noindent Let $q_1,q_2,...,q_N$ be the coordinates of $N$
particles on
the circle, interacting with the integrable potential $\sum
_{j<k}^N\wp(q_j-q_k)$, where $\wp$ is the Weierstrass elliptic function.
We
show that every symmetric elliptic function in $q_1,q_2,...,q_N$ is a
meromorphic function in time. We give explicit formulae for these
functions
in terms of genus $N-1$ theta functions.} \medskip\medskip\medskip

\section{Introduction}
\setcounter{equation}{0}

The elliptic Calogero system [Ca 1975]
\begin{equation}
\label{calogero}
\frac{d^2}{dt^2} q_i = - \sum_{j\neq i} \wp'(q_i-q_j), \;\; i=1,2,...,N
\end{equation} is a canonical Hamiltonian system, describing the motion
of $N$
particles on the circle $S^1= {\Bbb R}/\omega {\Bbb Z}$, $\omega \in
{\Bbb
R}$, with Hamiltonian (energy)
\be H=\frac{1}{2}\,\sum _{j=1}^Np_j^2+\sum _{j<k}^N\wp(q_j-q_k),\ee
where
$\wp(q)=\wp(q\vert \omega ,\omega ')$ is the Weierstrass elliptic
function
\be
\label{wpf}
\wp(q\vert \omega ,\omega ')=\sum _{m,n \in \Bbb Z}(q+m\omega +n\omega
')^{-2},\; \omega '/\omega \not \in {\Bbb R}.
\ee Denote by $\Gamma_1$ the elliptic curve
${\Bbb C}/ \{{2\omega \Bbb Z}+ 2\omega ' {\Bbb Z} \}$  with period
lattice
generated by
$2\omega$ and $2\omega'$. The Hamiltonian $H$ is invariant under the
obvious
action of the permutation group ${\cal S}_n$, so the phase space of the
compexified system is the cotangent bundle $T^*(S^N\Gamma_1)$ of the
$N$th
symmetric product $S^N \Gamma _1$.

It is known that this system has two Lax representations ( [Ca 1975],
[Kr
1980], see also [Pe 1990] for details ). The Lax operator $L$ defines
$N$
integrals of motion $I_k (p,q) =k^{-1} {\rm tr} (L^k), k=1, ..., N$. It
was
proved in [Pe 1977] that these integrals are in involution and hence
this
system is completely integrable in the Jacobi--Liouville sense [Ja
1842/43],
[Li 1855].

The Krichever Lax pair has a spectral parameter. This means that the
equations of motion of the system under consideration are equivalent to
the
matrix equation
\be
\label{lax} i\,\dot L(\lambda )=[L(\lambda ),M(\lambda )],
\ee where $L(\lambda )=L(p,q;\,\lambda )$ and $M(\lambda
)=M(p,q;\,\lambda )$
are two matrices of order $N$:
\begin{eqnarray}
\label{LL}
\{L(\lambda )\}_{jk}&=&p_j\,\delta _{jk}+i\,(1-\delta _{jk})\, \Phi
(q_j-q_k,\,\lambda );\\
\{M(\lambda )\}_{jk}&=&\delta _{jk}\,\left(\sum _{l\neq j}\wp(q_j-q_l)-
\wp(\lambda )\right)\nonumber \\ &&+\,(1-\delta _{jk})\,\Phi
'(q_j-q_k,\,\lambda );\\ \Phi (q,\lambda )&=&\frac{\sigma (q-\lambda
)}{\sigma (q)\,\sigma (\lambda )} \,\exp \left(\zeta (\lambda
)\,q\right);\\
\sigma (q)&=&q\,{\prod _{m,n}}^{\prime}
\left( 1-\frac{q}{\omega _{mn}}\right) \exp\left[ \frac{q}{\omega
_{mn}}+
\frac12\left( \frac{q}{\omega _{mn}}\right) ^2\right] , \end{eqnarray}
\[
\zeta (q)=\frac{\sigma '(q)}{\sigma (q)},\quad \omega _{mn}= m\omega
+n\omega
'.\] As it was shown by Krichever [Kr 1980], the equations of motion may
be
``linearized" on the Jacobian of the spectral curve \be
\label{kr}
\Gamma^N = \{ (\lambda ,\mu ): f(\lambda ,\mu )\equiv
\mbox{det}\,(L(\lambda
)-\mu I)=0 \} .\ee Namely, let
\begin{equation}
\theta({\bf z} \vert B) =
\sum_{{\bf N}\in {\Bbb Z}^N}e^{\pi i\langle {\bf N},B{\bf N}\rangle
+2\pi i
\langle {\bf N,z}\rangle} , \; {\bf z} \in {\Bbb C}^N \label{thk}
\end{equation} be the Riemann theta function with period matrix $B$,
where
$$B= (B_{ij}),\quad B= B^t,\quad {\rm Im}\, B > 0,\quad \langle
x,y\rangle =
\sum_j x_jy_j \; ,\; i,j = 1,...,N .
$$ It has been shown by Krichever [Kr 1980] that, if $B$ is the period
matrix
of the curve
$\Gamma^N$, then for suitable constant vectors $U,V,W \in {\Bbb C}^N$
and for
a fixed parameter $t\in {\Bbb C}$, the equation \begin{equation}
\theta (Uq+Vt+W)= 0, \; q\in \Bbb C
\label{sol1}
\end{equation} has exactly $N$ solutions $q = q_j(t)$ on the Jacobian
Jac
$(\Gamma ^N)$ of the curve $\Gamma ^N$. The functions $q_j(t)$ provide
solutions of the elliptic Calogero system (\ref{calogero}). The equation
(\ref{sol1}) for these solutions is, however, not explicit and seems to
be
not well understood.

The aim of the present paper is to give ``the effectivization" of these
formulae based on the projection method by Olshanetsky and Perelomov
([OP
1976], [OP 1977]) of explicit integration of the equations of motion in
the
rational and the trigonometric cases, as well  on the algebro-geometric
approach of Krichever [Kr 1978], [Kr 1980].

\section{Explicit solutions}

Let $\Gamma_N$ be a genus $N$ Riemann surface which is an $N$-sheeted
covering of an elliptic curve $\Gamma_1$ \begin{equation}
\Gamma_N \stackrel{\pi}{\rightarrow} \Gamma_1 \; \; . \end{equation} It
follows from a theorem of Weierstrass (see for example [Ko 1884], [Po
1884],
[Po 1886] and [BBEIM 1994], Theorem7.4) that the period matrix of the
curve
$\Gamma _N$ in a suitable basis has the form $(I,B)$ where $I={\rm diag}
(1,1,...,1)$, and \begin{equation} B=\left(\begin{array}{ccccc}
\frac{\tau}{N} &\frac{k}{N}&0&\ldots &0\\ \frac{k}{N}&b _{22}&b
_{23}&\ldots
&b _{2N}\\ 0&b _{32}&b _{33}&\ldots &b _{3N}\\
\vdots &\vdots &\vdots &\vdots&\vdots \\ 0 &b _{N2}&b _{N3}&\ldots &b
_{NN}
\end{array}\right) \,.
\end{equation} for a suitable positive integer $k$. Consider the Riemann
theta function $\theta(x,{\bf t}) = \theta(x,{\bf t}\vert B)$, where
${\bf t}=(t_1,t_2,...,t_{N-1})$, $(x,{\bf t})\in {\Bbb C}^N$. We have

\begin{equation}
\theta(x+1,{\bf t})=\theta(x,{\bf t}),
\theta(x+\tau ,{\bf t})= e^{-2\pi i N x - \pi i N \tau } \theta(x,{\bf
t}),
i= \sqrt{-1}
\label{theta1}
\end{equation} and therefore for any fixed ${\bf t}$ the function
$\theta(x,{\bf t})$ is an elliptic theta function of order $N$ [Du
1981]. In
particular it has exactly $N$ zeros on $\Gamma _1= {\Bbb C}/ \{{\Bbb Z}+
\tau
{\Bbb Z} \}$ which we denote by $x_i({\bf t})$, $i=1,2,...,N$.
\begin{lemma}.
\label{prk} The following identity holds
$$
\frac{\partial^2}{\partial x^2} \log \theta(x,{\bf t}\vert B)=
\sum_{i=1}^N
\wp(x- x_i({\bf t}) \vert \tau ) +N \frac{\theta_1'''(0)}{3\theta_1'(0)}
\; ,
$$ where {\rm [BE 1955]}
$$\theta _1(x\vert \tau ) = \theta
\,\left[\begin{array}{c}1/2\cr 1/2\end{array} \right]\,(x\vert \tau ) \;
$$
\end{lemma} {\bf Proof}. The relations
\begin{equation}
\label{sigma1}
\theta _1(x+1)=-\theta _1(x), \theta _1(x+\tau )=- e^{-2\pi ix-\pi i\tau
}\theta _1(x)
\end{equation} compared to (\ref{theta1}) imply that
\begin{equation}
\left( \frac{\theta(x,{\bf t})} { \prod_{i=1}^N \theta _1(x- x_i({\bf
t})
)}\right)^2 \label{th2}
\end{equation} is a meromorphic function in $x$ on $\Gamma _1$ which has
no
poles, and hence it is a constant (in $x$). It follows that $$
\frac{\partial^2}{\partial x^2} \log
\frac{\theta(x,{\bf t})} { \prod_{i=1}^N \theta _1(x- x_i({\bf t}) ) }
\equiv
0 \; . $$ Finally we use that
\be
\wp(x)= - \frac{\partial^2}{\partial x^2} \log \sigma (x),\quad \theta
_1(x)=c \exp(\eta x^2)\,\sigma (x) \ee where
$$
\eta = -\frac{\theta_1'''(0)}{6\theta_1'(0)} $$ and $c$ is a suitable
constant [BE 1955]. $\Box$

\begin{theorem}.
\label{period} The Krichever curve $\Gamma ^N$ is an $N$-sheeted
covering of
an elliptic curve $\Gamma _1 = {\Bbb C}/
\{ {\Bbb Z} +
\tau {\Bbb Z} \}$. There exists a canonical homology basis and a
normalized
basis of holomorphic one-forms on $\Gamma ^N$, such that the
corresponding
period matrix of
$\Gamma _N$ takes the form $(I,B)$, where $I={\rm diag}(1,1,...,1)$, and

\begin{equation}
\label{rb} B=\left(\begin{array}{ccccc}
\frac{\tau}{N} &\frac{1}{N}&0&\ldots &0\\ \frac{1}{N}&b _{22}&b
_{23}&\ldots
&b _{2N}\\ 0&b _{32}&b _{33}&\ldots &b _{3N}\\
\vdots &\vdots &\vdots &\vdots&\vdots \\ 0 &b _{N2}&b _{N3}&\ldots &b
_{NN}
\end{array}\right) \,.
\end{equation}

In the same basis the vectors $U$ and $V$ in (\ref{sol1}) read

\begin{equation} U= (1 ,0,...,0), V=(0,V_2,...,V_N)\; .
\label{equv}
\end{equation}

\end{theorem}
A direct proof (without using the Weierstrass theorem) of the
above Theorem will be given in the last section. From now on we make the
convention that
$2\omega =1$ so the period lattice of $\Gamma _1$ is
 $${\Bbb Z} +
\tau {\Bbb Z}, \quad \tau = 2\omega '/2\omega = 2\omega ' \; .
$$
\begin{corollary}.
The symmetric functions
$$f_k(t)= \sum_{i=1}^N \wp^{(k)}(q_i(t))$$
 are
meromorphic in $t$. Explicit formulae for them are obtained from Lemma
\ref{prk}:
$$ f_0(t)=
\frac{\partial^2}{\partial x^2} \log \theta(x,{\bf t})\vert _{x=0} -N
\frac{\theta_1'''(0)}{3\theta_1'(0)}
$$
$$ f_k(t)=(-1)^k
\frac{\partial^{k+2}}{\partial x^{k+2}} \log \theta(x,{\bf t})\vert
_{x=0},
\; k>0,
$$ where
$$ {\bf t}= (V_2t+W_2,V_3t+W_3,...,V_N t+W_N) \; . $$
\end{corollary}

Our next construction is motivated by [OP 1976], [OP 1977] and [Kr
1980]. Let
us define the function
\be
\label{fxt1} F(x,t)=
\prod _{j=1}^N\frac{
\sigma (x-q_j(t))} {\sigma(x) \sigma(q_j(t))}= [\theta_1'(0)]^{-N}]
\prod _{j=1}^N\frac{
\theta _1(x-q_j(t))} {\theta_1(x) \theta_1(q_j(t))} ,\quad \sum
_{j=1}^Nq_j(t)=0
\ee where
$$ q_j(t), \quad t\in {\Bbb C},\quad j= 1,2,...,N,
$$ is a solution of the elliptic Calogero system. \begin{lemma}.
\label{fmer}
$F(x,t)$ is a meromorphic function in $x$ on $\Gamma _1$ and meromorphic
function in $t$ on ${\Bbb C}$, explicitly given by \begin{equation}
\label{fxt2} F(x,t)= [-\theta_1'(0)]^{-N}] \frac{
\theta(Ux+Vt+W)}{\theta_1(x)^N \theta(Vt+W)} \; .
\end{equation}
\end{lemma} {\bf Proof}. We already noted that the function (\ref{th2})
is a
constant in $x$, and hence $$
\frac{\theta(x,{\bf t})} { \prod_{i=1}^N \theta _1(x- x_i({\bf t}) )}
\equiv
\frac{\theta(0,{\bf t})} { \prod_{i=1}^N \theta _1(- x_i({\bf t}) )} \;
.
$$ This combined with (\ref{equv}) gives
$$
\frac { \prod_{i=1}^N \theta _1(x- q_i(t) )} { \prod_{i=1}^N \theta _1(
q_i(t) )}= (-1)^N \frac{\theta(Ux+Vt+W)}{\theta (Vt+W)} \; . $$
$\Box$

The expansion of
$F(x,t)$ on the basis of first order theta functions in $x$ defines
$(N-1)$
meromorphic functions in the variables
$q_1,\ldots ,q_N$ which are also meromorphic functions in $t$ with only
simple poles. Hence we can take them as new ``good" variables. The
expansion
of $F(x,t)$ can be obtained by making use of the addition formulae for
elliptic functions. In the case $N=2$, we have the following ``addition
formula" [BE 1955] \be F(x,t)= - \frac{\sigma (x-q)\,\sigma (x+q)}
{\sigma
^2(x)\sigma ^2(q)} = \wp(x) - \wp(q) \; ,\ee which generalizes for
arbitrary
$N$ in the following way
\begin{lemma}.
\label{fx1} For any ${\bf q}=(q_1,q_2,...,q_N),x$, such that $\sum
q_j=0$
define \begin{equation} F(x,{\bf q})= \prod _{j=1}^N \frac{\sigma
(x-q_j)}
{\sigma(x) \sigma(q_j)}
\end{equation}
\begin{equation}
\Delta ({\bf q})= (N-1)!
\det \left\vert
\begin{array}{ccccc} 1 &\wp(q_1) &\wp'(q_1)&\ldots &\wp^{(N-3)}(q_1)\cr
1
&\wp(q_2) &\wp'(q_2)&\ldots &\wp^{(N-3)}(q_2)\cr
\ldots&\ldots&\ldots&\ldots&\ldots \cr 1 &\wp(q_{N-1})
&\wp'(q_{N-1})&\ldots
&\wp^{(N-3)}(q_{N-1}) \end{array}
\right \vert \; .
\end{equation} The following identity holds
\begin{equation}
\label{ad1} F(x,{\bf q}) \Delta ({\bf q}) \equiv
\det \left\vert
\begin{array}{ccccc} 1 &\wp(x) &\wp'(x)&\ldots &\wp^{(N-2)}(x)\cr 1
&\wp(q_1)
&\wp'(q_1)&\ldots &\wp^{(N-2)}(q_1)\cr
\ldots&\ldots&\ldots&\ldots&\ldots \cr
1 &\wp(q_{N-1}) &\wp'(q_{N-1})&\ldots &\wp^{(N-2)}(q_{N-1}) \end{array}
\right\vert
\end{equation}
\end{lemma} {\bf Remark}. The substitution $x=q_N$ in (\ref{ad1}) gives
the
following addition formula for the Weierstrass $\wp$-function
\begin{equation}
\label{ad2}
\det \left\vert
\begin{array}{ccccc} 1 &\wp(q_1) &\wp'(q_1)&\ldots &\wp^{(N-2)}(q_1)\cr
1
&\wp(q_{2}) &\wp'(q_{2})&\ldots &\wp^{(N-2)}(q_{2})\cr
\ldots&\ldots&\ldots&\ldots&\ldots \cr 1 &\wp(q_N) &\wp'(q_N)&\ldots
&\wp^{(N-2)}(q_N) \end{array}
\right\vert \equiv 0 \; .
\end{equation} {\bf Proof}. For fixed ${\bf q} = (q_1,q_2,...,q_N)$ the
functions in the left and right-hand side of the identity (\ref{ad1})
are
meromorphic in $x$ on the elliptic curve
$\Gamma _1$. Both of them have a pole of order $N$ at $x=0$ and simple
zeros
at
$x=q_1,...,q_{N-1}$. It follows that their ratio is a first order
elliptic
function, and hence a constant in $x$. To compute this constant we use
that
$\sigma (x)=x+...$, $\wp(x)= 1/x^2+...$, and then compare the Laurent
series
of the two functions in a neighborhood of $x=0$. $\Box$

Note finally that if for fixed ${\bf q}$ and $\tilde {\bf q}$ holds
$F(x,{\bf q})\equiv F(x,\tilde {\bf q})$, then up to a permutation ${\bf
q}=\tilde{\bf q}$. Therefore there is a one-to-one correspondence
between the
coefficients of $\wp^{k}(x)$ in the expansion of $F(x,{\bf q})$, and the
points of the $(N-1)$th symmetric power of the elliptic curve $\Gamma _1
\backslash \{ 0 \} $. In particular every meromorphic function on this
symmetric power is a rational function in the above coefficients. This
implies the following \begin{corollary}. Let $f(x)$ be a meromorphic
function
on the elliptic curve $\Gamma _1$, and let $S$ be a symmetric rational
function in $N-1$ variables. If $q_1(t),q_2(t),...,q_{N}(t)$, $\sum
q_i\equiv
0$ is a solution of the elliptic Calogero system, then
$S(f(q_1(t),f(q_2(t)),...,f_{N-1}(q_{N-1}(t)))$ is a meromorphic
function in
$t$.
\end{corollary}

The further analysis of the explicit formulae for the solutions of the
elliptic Calogero system can be based on Lemma \ref{fmer}, Lemma
\ref{fx1},
and the identity
$$F(x,t) \equiv F(x, {\bf q}(t)) \; .$$

Consider the seemingly trivial case of two particles ($N=2$). Let us
give
first an explanation of the Krichever formula (\ref{sol1}) for the
solutions
$q_1(t)=-q_2(t)$. Put $q_1-q_2=q$ and $p_1=-p_2=p$. The Hamiltonian $H$
becomes $H(p,q)= p^2+ \wp(q)$, and the reduced Hamiltonian system is
\be
\label{ham}
\frac{d}{dt}q=2p, \frac{d}{dt}p=-\, \wp'(q) ,\; (q,p)\in T^*\Gamma _1\; .
\ee
The Lax matrix $L$ is
$$ L(\lambda)=
\left(
\begin{array}{lr} p& i \Phi (q,\lambda)\\ i\Phi (-q,\lambda)&-p
\end{array}\right) \;
$$ and the corresponding spectral polynomial $$ {\rm det}(L(\lambda)-\mu
I)=
\mu^2 -p^2+ \Phi(q,\lambda)\Phi(-q,\lambda) $$
$$ =
\mu^2-p^2+ \wp(\lambda) - \wp(q)=
\mu^2+ \wp(\lambda)-H(p,q) \;
$$ defines a spectral curve
$$
\Gamma _2= \{(\mu ,\lambda ): \mu ^2+ \wp(\lambda )=h\} \; . $$ Suppose
that
$h=H(p,q)$ is fixed in such a way, that the meromorphic function $\wp
(\lambda )-h$ has two distinct zeros on $\Gamma _1$. The spectral curve
$\Gamma _2$ is a double ramified covering over the elliptic curve
$\Gamma _1$
with projection map
$\pi :\Gamma _2\rightarrow\Gamma _1: (\mu ,\lambda) \rightarrow \lambda
$. It
follows that $\Gamma _2$ is a genus two curve with holomorphic
differentials
$$
\omega_1= d \lambda,\quad \omega_2= \frac{d \lambda }{\mu} \; . $$
On the other hand $\Gamma _2$ is identified to the orbit
$$
\{(p,q)\in T^*\Gamma _1= : H(p,q)=h \}
$$
under
the map
$$
(p,q) \rightarrow (\mu ,\lambda )\; .
$$
Consider further the embedding of the orbit $\Gamma _2$ into its Jacobian
variety $Jac(\Gamma _2)$
\begin{equation}
\label{t1}
\Gamma _2 \rightarrow Jac(\Gamma _2) :
P\rightarrow (\int_{P_0}^P d\lambda ,
\int_{P_0}^P \frac{d\lambda }{\mu }) \; .
\end{equation}
By the Riemann theorem [GH, 1978],
the curve $\Gamma _2 \subset Jac(\Gamma _2)$ defines a divisor which coincides,
up to addition of a constant, with the
 Riemann theta divisor $\Theta  \subset  Jac(\Gamma _2)$ on the Jacobian
variety $Jac(\Gamma _2)$.

Let $(p(t),q(t))$ be a solution of the elliptic Calogero system, with initial
condition $(p(t_0),q(t_0))= P_0$.
Taking into consideration that
$$
\frac{d\lambda }{\mu }= 2\, dt,  \quad
d \, \lambda = dq\; ,\quad (\lambda ,\mu )\in \Gamma _2
$$
the formula (\ref{t1}) takes the form
\begin{equation}
\label{t11}
 T^* \Gamma _1 = {\Bbb C} \times \Gamma _1 \ni
(p(t),q(t)) \rightarrow (2t-2t_0, q(t)-q(t_0)) \in Jac(\Gamma _2) \; .
\end{equation}
It follows that there exist constant
vectors
${\bf a,b,c} \in \C^2 $ such that
\be
\label{t3}
\theta ({\bf a} q(t) + {\bf b} t + {\bf c} ) \equiv 0 \; .
\ee
Of course these constants depend  on the choise of symplectic homology basis
and the choice of normalized basis  of holomorphic one-forms.  Namely, let
$a,b$ be two loops on $\Gamma _1$, such
that $\pi^{-1}(a)=\{a_1,a_2\}$, $\pi^{-1}(b)=\{ b_1,b_2 \}$, where
$a_i,b_j$
represent an integer symplectic homology basis on $\Gamma_2$: $a_i\circ
b_j=\delta _{ij}$,
$a_i\circ a_j=0$, $b_i\circ b_j=0$. Then $$
\int_{a_1} d\lambda = \int_{a_2} d\lambda,\; \; \int_{b_1} d\lambda =
\int_{b_2} d\lambda $$
$$
\int_{a_1} \frac{d \lambda}{\mu }= - \int_{a_2} \frac{d \lambda}{\mu
},\;\;
\int_{b_1} \frac{d \lambda}{\mu }= - \int_{b_2} \frac{d \lambda}{\mu }
\; \;
. $$ If we define a new symplectic basis
$$
\tilde{a}_1=a_1+a_2, \tilde{a}_2= b_1-b_2, \tilde{b}_1=b_1,
\tilde{b}_2=a_2 $$
and normalize the two holomorphic one-forms as $$ d\lambda \rightarrow
\frac{d\lambda }{\int_{\tilde{a}}\pi ^* d\lambda } = \frac{d\lambda
}{2\int_a
d\lambda },\;
\frac{d\lambda }{\mu }\rightarrow
\frac{{d\lambda}/{\mu} }{ \int_{\tilde{a}_2} {d\lambda }/{\mu }},
$$ then the period matrix of $\Gamma _2$ takes the form $$
\left(\begin{array}{cccc} 1 &0&\tau_1/2 &1/2 \cr 0&1&1/2&\tau_2/2
\end{array}
\right)
$$ where
$$
\tau_1 = \frac{\int_b d\lambda }{\int_a d\lambda }, \tau_2= \frac{
\int_{a_2}{d\lambda }/{\mu }}{ \int_{b_1} {d\lambda }/{\mu }} \;\;\;.
$$
This, together with (\ref{t11}) implies that
$$
{\bf
a}=(\frac{1}{\int_{\tilde{a}_1} d\lambda}, 0)= (\frac{1}{2\int_{a}
d\lambda},
0) , {\bf b}= (0, \frac{1}{\int_{b_1}{d\lambda }/{\mu }}) \;\;\; .
$$
Finally the vector ${\bf c}$ is arbitrary and plays the role of initial
condition. The function $F(x,t)$ defined in (\ref{fxt1}) takes the form
\be
F(x,t)=- \frac{\sigma (x-q(t))\,\sigma (x+q(t))} {\sigma ^2(x)\sigma
^2(q(t))}
\ee and hence [BE,1955], [WW 1927]
\be F(x,t)= \wp(x) - \wp(t) \; .
\ee So  the elliptic function $\wp(q\vert \omega ,\omega ')$, and also
\be
\mbox{sn}^2(q, k)\sim \frac{\theta _1^2(q\vert k)}{\theta _4^2 (q\vert
k)},\quad \mbox{cn}^2(q,k)\sim
\frac{\theta _2^2(q\vert k)} {\theta _4^2(q\vert k)},\quad
\mbox{dn}^2(q,k)\sim \frac{\theta _3^2 (q\vert k)}{\theta _4^2(q\vert
k)}\ee
are ``good" variables (in the sense that they are meromorphic in $t$).
The
equation of motion for them takes a very simple form. We get
\be \mbox{sn}^2(q,k)=1-a^2+a^2\,\mbox{sn}^2(\gamma t, \tilde k),\ee
where
\begin{equation} a^2= \frac{h-1}{h},\, \gamma = 2(h-k^2),\, \tilde k^2 =
\frac{h-1}{h-k^2}\, k^2 \; . \end{equation} One can easily show that the
even
functions $\mbox{cn}\,(q,k)$ and $\mbox{dn}\,(q,k)$ (but not
$\mbox{sn}\,(q,k)$) are ``good" variables and we get as in [Pe 1990]
\begin{eqnarray}
\mbox{cn}\,(q,k)&=&\alpha \,\mbox{cn}\,(\gamma t,\tilde k),
\label{e11}
\\
\mbox{dn}\,(q,k)&=&\beta \,\mbox{dn}\,(\gamma t,\tilde k),\, b= (k/
\tilde k)a
.
\label{e12}
\end{eqnarray}

\section{Reduction of theta functions} \setcounter{equation}{0} The
reduction
theory was elaborated by Weierstrass (see for example [Ko 1884]) and
Poincar\'e [Po 1884], [Po 1886]. Consider first the case $N=2$. The
Riemann
theta function associated with the Riemann matrix (\ref{rb}) has the
form:
\be
\theta (z_1,z_2)=\sum _{n_i,n_j}\exp\{ i\pi \,[B_{ij}n_in_j+2n_jz_j]\},
\quad
i, j=1,2.
\ee where
$$ B_{11}=\tau _1/2,\quad B_{22}=\tau _2/2,\quad B_{12}= B_{21}=1/2. $$
A
straightforward computation gives
$$
\theta(z_1,z_2) =
\sum _{n_1,n_2}\exp\{ i\pi \, [\tau _1 \frac{n_1^2}{2}+n_1n_2 +\tau
_2\frac{n_2^2}{2}+2n_1z_1+2n_2z_2 ] \}
$$
$$ = \sum _{k_1,n_2\in {\Bbb Z}}\exp\{ i\pi \, [2\tau _1 k_1^2+
4k_1z_1]\}
\exp\{ i\pi \, [\tau _2 \frac{n_2^2}{2} + 2n_2z_2]\} $$
$$ +
\sum _{k_1,n_2\in {\Bbb Z}}\exp\{ i\pi \, [2\tau _1(k_1+\frac{1}{2})^2+
4(k_1+\frac{1}{2})z_1]\}
\exp\{ i\pi \, [\tau _2 \frac{n_2^2}{2} + 2(n_2+\frac{1}{2})z_2]\}$$ $$
=\theta _3(2z_1\vert 2\tau _1)\, \theta _3(z_2\vert \frac{\tau
_2}{2})+\theta
_2(2z_1\vert 2\tau _1)\,\theta _4(z_2\vert \frac{\tau _2}{2}) $$ where
$\theta _1, \theta _2, \theta _3$ and $\theta _4$ are defined by
formulae:
\begin{eqnarray}
\theta _1(z\vert \tau )&=&\theta \left[\begin{array}{c}1/2\\ 1/2
\end{array}\right] (z\vert \tau )=2q^{1/4}\sum _{n=1}^\infty (-1)^n
q^{n(n+1)}\sin[(2n+1)\pi z];\\
\theta _2(z\vert \tau )&=&\theta \left[\begin{array}{c}1/2\\ 0
\end{array}\right] (z\vert \tau )=2q^{1/4}\sum _{n=1}^\infty
q^{n(n+1)}\cos
\, [(2n+1)\pi z];\\
\theta _3(z\vert \tau )&=&\theta \left[ \begin{array}{c}0\\ 0\end{array}
\right] (z\vert \tau)=1+2\sum _{n=1}^\infty q^{n^2}\cos (2\pi nz);\,\,
q=\exp(i\pi \tau );\\
\theta _4(z\vert \tau )&=&\theta \left[ \begin{array}{c}0\cr
1/2\end{array}
\right](z,\tau )=1+2\sum _{n=1}^\infty (-1)^n\,q^{n^2}\,\cos\,(2\pi nz).
\end{eqnarray} So in this case, the equation $\theta(z_1,z_2)=0 $ is
equivalent either to \be A\,\mbox{dn}\,(2z_1\vert 4\tau
_1)\,\mbox{dn}\,(z_2\vert \tau _2)+ \mbox{cn}\,(2z_1\vert 4\tau
_1)=0,\ee or
to \be A\,\mbox{dn}\,(2z_2\vert 4\tau _2)\,\mbox{dn}\,(z_1\vert \tau
_1)+
\mbox{cn}\,(2z_2\vert 4\tau _2)=0,
\ee where
\be A=\frac{\theta _3(0\vert 4\tau _1)\,\theta _3(0\vert \tau
_2)}{\theta _2
(0\vert 4\tau _1)\,\theta _4(0\vert \tau _2)}\ee or \be
\mbox{dn}\,(z_1\vert
\tau _1)=B\,\mbox{dn}\,(2iz_2+K\vert \tilde \tau _2). \ee Let us give
also a
more symmetric form of the theta divisor for this case: $$
\mbox{dn}(2z_1,k_1)\,\mbox{dn}(2z_2,k_2)+\mbox{dn}(2z_1,k_1)\,
\mbox{cn}(2z_2,k_2)+\mbox{cn}(2z_1,k_1)\,\mbox{dn}(2z_2,k_2) $$ \be
\hfil -\, \mbox{cn}(2z_1,k_1)\,\mbox{cn}(2z_2,k_2)=0. \ee

Using the constraint $\theta ({\bf a}x+{\bf b}t+{\bf c})=0$ and taking
$z_1=q$,
$z_2 = (1/2) K+i\gamma t$, we get once again (\ref{e11}), (\ref{e12}).

Consider now the case of arbitrary $N$. Let $\theta(z_1,z_2,...,z_N
\vert B)$
be the Riemann theta function with period matrix as in Theorem
\ref{period}.
In a quite similar way we get \be
\label{eqe}
\theta (z_1,z_2,\ldots ,z_N)=\sum _{j=0}^{N-1}\theta _j(z_1)\, \Theta
_j(z_2,\ldots ,z_N),
\ee where
\be \theta _j(z_1)=\theta \left[ \begin{array}{c}
j/N\\0\end{array}\right]
(Nz_1\vert N^2\tau _1),
\ee
\be
\Theta _j(z_2,\ldots , z_N)=\Theta \left[ \begin{array}{cccc} 0&0&\ldots
&0\\
j/N&0&\ldots &0\end{array}\right] (z_2,\ldots ,z_N\vert \hat B). \ee In
the
above formula
$ \hat B$ is the right lower $(N-1)\times(N-1)$ minor of $B$ (\ref{rb}),
and
the theta functions with fractional characteristics are defined for
example
in [Kr 1903], [Ko 1976], [Du 1981], [BBEIM 1994]. A reduction formula
similar
to (\ref{eqe}), but containing $N^2$ terms, can be found in [BBEIM
1994],Corollary 7.3.

\section{Geometry of the spectral curve} \setcounter{equation}{0}
In this
section we prove Theorem \ref{period}.

Let $\Gamma_N$ be a genus $N$ Riemann surface which is an $N$-sheeted
covering of an elliptic curve $\Gamma_1$ \begin{equation}
\label{covering}
\Gamma_N \stackrel{\pi}{\rightarrow} \Gamma_1 \; \; . \end{equation}
Choose
two loops $a,b $ which generate the fundamental group
$\pi_1(\Gamma_1,P)$,
$P\in\Gamma_1$, and denote
$\check{\Gamma }_1= \Gamma_1 \backslash \{a\cup b\}$.
 Let us suppose for simplicity
that the ramification points of the projection map $\pi$ are distinct.
Connect further these ramification points by non-intersecting arcs
$\gamma_i
\subset
\check{\Gamma }_1$. The set $\pi^{-1}(\check{\Gamma }_1 \backslash
\cup_i
\gamma_i)$ is a disjoint union of $N$ ``sheets". To reconstruct the
topological covering (\ref{covering}) we have to indicate how the
opposite
borders of the cuts $\gamma_i$ are glued, as well how the opposite
borders of
the (pre-images of the) cuts $a$ and $b$ respectively are glued
together.
Thus there is only a finite number of topologically different coverings
(\ref{covering}). It may be shown that the Krichever curve (\ref{kr}) is
of
genus at most $N$,
 and for generic $(p_i,q_i)$ its genus is exactly $N$. The projection
map $
\pi $ (\ref{covering}) is defined then by $\pi (\mu ,\lambda )= \lambda
$.
>From now on we shall always assume that $(p_i,q_i)$ are generic. In the
case
when
$\Gamma_N$ is the genus $N$ Krichever spectral curve (\ref{kr}), and
$\Gamma_1$ is the elliptic curve with half periods $\omega, \omega'$,
the
covering (\ref{covering}) has a number of special properties.

To prove (\ref{rb}) we shall need the following

\begin{proposition}.
\label{connected} Let $\Gamma_N$ be the Krichever curve (\ref{kr}).
There
exist loops $a,b \in \pi_1(\Gamma_1,P)$ such that, if $\check{\Gamma
}_1=\Gamma_1 \backslash \{a\cup b\}$, $\partial \check{\Gamma }_1=
a\circ
b\circ a^{-1} \circ b^{-1}$, then

\noindent {\bf i.} $\pi^{-1}(\check{\Gamma }_1)$ is connected

\noindent {\bf ii.} $\pi^{-1}(\partial \check{\Gamma }_1)$ has exactly
$N$
connected components.
\end{proposition}

On its hand the above proposition implies the following
\begin{proposition}.
\label{basis} There exist loops $a,b \in \pi_1(\Gamma_1,P)$, $P\in
\Gamma_1$,
such that $$\pi^{-1}(a)= \{a_1,a_2,...,a_N\}, \pi^{-1}(b)=
\{b_1,b_2,...,b_N\}$$ where $a_i,b_i$ represent a symplectic homology
basis
of $H_1(\Gamma_N,\Z)$, $a_i\circ b_j= \delta_{ij}$.
\end{proposition}

\noindent {\bf Proof of (\ref{rb}) assuming Proposition \ref{basis}}.

\noindent Let $d \lambda$ be the holomorphic one-form on $\Gamma_1$.
Then the
pullback $\pi^* d \lambda$ of $d \lambda$ is a holomorphic one-form on
$\Gamma_N$ and we have
$$
\int_{a_i} \pi^*d \lambda =\int_{a} d \lambda,
\int_{b_i} \pi^*d \lambda =\int_{b} d \lambda \; . $$ Choose the
following
new integer homology basis of $\Gamma_N$ $$
\tilde{a}_1= a_1+a_2+...a_N,\quad \tilde{b}_1=b_1, $$
$$
\tilde{a}_2= N b_1-b_1-b_2-...-b_N, \quad \tilde{b}_2 = a_2 $$ and
$$
\tilde{a}_i= b_i-b_1,\quad \tilde{b_i}= a_2-a_i, \quad i=3,...,N \; . $$
This
is also a symplectic basis of $H_1(\Gamma_N,\Z)$, as $$
\sum _{i=1}^N \tilde{a}_i \wedge \tilde{b_i}= \sum _{i=1}^N a_i \wedge
b_i \;
.
$$ Let $\omega_1, \omega_2,..., \omega_N$ be a basis of holomorphic
one-forms
on $\Gamma_N$, such that
$$
\omega_1= \frac{d \lambda}{\int_{\tilde{a}_1} d \lambda} ,\,
\int_{\tilde{a}_i} \omega_j = \delta_{ij} \; . $$ Then $B=
(\int_{\tilde{b}_j} \omega_i)_{i,j}^{N,N}$ is a symmetric matrix with
positive definite imaginary part, such that
$$
\int_{\tilde{b}_1} \omega_1 = \frac{\tau}{N},\, \int_{\tilde{b}_2}
\omega_1 =
\frac{1}{N},\, \int_{\tilde{b}_i} \omega_1 =0, i\geq 3 \; $$ which
completes
the proof of  \ref{rb}. \medskip

\noindent {\bf Proof of Proposition \ref{connected}}.

\noindent First of all let us note that if the claim holds for some
Krichever
curve, then it holds for any Krichever curve. Indeed, the space of all
such
curves is parameterized by $\C^{N-1}$ (the first integrals of the
integrable
Hamiltonian system (\ref{lax})) and hence it is connected. Let us fix a
generic point $(p_i,q_i)$, $i=1,2,...,N$. It is enough to prove now our
proposition for at least one pair of half-periods $\omega, \omega'$, for
example for $\vert \omega \vert, \vert \omega' \vert \sim \infty$.

Let us represent
$
\check{\Gamma }_1 \subset \C=\bbbp^1 \backslash \infty
$ as the interior of the period parallelogram formed by $2\omega$ and
$2\omega'$. When
$\vert \omega \vert \rightarrow \infty$, $\vert \omega' \vert
\rightarrow
\infty$, the boundary. $\partial \check{\Gamma }_1= a\circ b\circ a^{-1}
\circ b^{-1}$ tends to $\infty \in \bbbp^1$, and
$\check{\Gamma }_1$ tends to $\check{\Gamma }^\infty_1= \C$.  In a
similar
way we define the ``limit" curve
$ \check{\Gamma }_N^\infty$ which is explicitly described in the
following
way. When
$\vert \omega \vert \rightarrow \infty$, $\vert \omega' \vert
\rightarrow
\infty$, then on any compact set the Weierstrass functions $\sigma(q),
\zeta(q), \wp(q)$ tend to $q, 1/q, 1/q^2$ respectively, and hence the
function $\Phi(q, \lambda)$ tends to
$$
\frac{q-\lambda}{q \lambda}\, \exp (q/\lambda ) . $$ Denote the
corresponding
``limit" Lax matrix (\ref{LL}) by $L^\infty(\lambda)$. The curve $
\check{\Gamma }_N^\infty$ is the affine curve $$
\{(\lambda,\mu): {\rm det} (L^\infty(\lambda) - \mu I_N)=0 \} $$
completed
with $N$ distinct points corresponding to $\lambda= 0$. The last holds
true
if and only if the ramification points of the projection map $\pi$
(\ref{covering}) tend to some values different from $\lambda=0$ (it is
easy
to check that this is a generic condition on $(p_i,q_i)$). We shall also
suppose that these values are different from $\lambda= \infty$ (another
generic condition). Under these restrictions one may prove (as in [Kr
1980])
that $ \check{\Gamma }_N^\infty$ is a Riemann sphere, with $N$ punctures
(the
pre-images of $\lambda= \infty$). We obtain thus a map $\pi :
\bbbp^1\rightarrow \bbbp^1$ with $2N-2$ ramification points different
from
$\lambda = 0, \infty$. The fact that $\pi^{-1}(\C)$ is connected implies
the
part {\bf i}. of the proposition, and the fact that $\pi^{-1}(\infty)$
is a
disjoint union of $N$ points implies {\bf ii}. \medskip

\noindent {\bf Proof of Proposition \ref{basis}}.

\noindent Let us represent
$\check{\Gamma }_N$ by a graph with $N$ vertices. A vertex corresponds
to a
sheet (see the beginning of this section), an edge connects two vertices
if
and only if the corresponding sheets have a common ramification point.
Proposition \ref{connected}. {\bf i}. implies that the graph is
connected,
and {\bf ii}. that each sheet contains an even number of ramification
points.
As the total number of ramification points is $2N-2$ and each point
belongs
to exactly two sheets, then in addition the graph of $\check{\Gamma }_N$
is
simply connected.

Consider now the punctured curve
$$\tilde{\Gamma}_1= \check{\Gamma}_1
\backslash \cup_i R_i\; ,
$$
 where $R_i, i=1,...,2N-2$ are the ramification points
of $\pi$. The fundamental group
$\pi_1(\tilde{\Gamma}_1,P)$ has a natural representation in the
permutation
group
$S_n$. Namely, when a point $Q\in \Gamma_1$ makes one turn along a loop
$a
\in \pi_1(\tilde{\Gamma}_1,P)$, the set $\pi^{-1}(P)= \cup_{i=1}^N P_i$
is
transformed to itself. If the loops $a$ and $b$ induce the identity
permutation, then $\pi^{-1}(a)$,$\pi^{-1}(b)$ are disjoint unions of $N$
loops with obvious intersections, which implies Proposition \ref{basis}.
If
not, we shall modify $a$ and $b$.

Let $c\in \pi_1(\tilde{\Gamma}_1,P)$ be a loop which makes one turn
around
some ramification point of $\pi $. Then $c$ induces a permutation which
exchanges the two sheets containing the ramification point. As the graph
of
$\check{\Gamma }_N$ is connected then all such transpositions generate
the
permutation group $S_n$. Thus for suitable $c$ the loop $a\circ c$
induces
the identity permutation. It remains to substitute $a\rightarrow a\circ
c$
and to note that $a= a\circ c$ in $ \pi _1(\Gamma _1,P)$. \medskip

\noindent {\bf Proof of (\ref{equv})} (compare to [Be 1994), Theorem
7.14).

\noindent Let $0\in \Gamma _1$ be the pole of $\wp(z)$. We denote $$
\pi ^{-1}(0)= \{ \infty_1, \infty_2, ..., \infty_N\},\quad \infty_i\in
\Gamma
_N \; .
$$ In a neighborhood of each point $\infty_i$ on the Krichever curve
$\{(\lambda ,\mu ): f(\lambda ,\mu) =0 \}$ the meromorphic function $\mu
$
has the following Laurent expansion [Kr 1978]
$$
\mu = -\frac1\lambda + O(1),\quad i=1,2,...,N-1, $$
$$
\mu = \frac{N-1}{\lambda }+O(1) \; .
$$ It follows that if
$$
\omega _j = f_j(P) d \lambda , \quad P=(\lambda ,\mu ) \in \Gamma_N $$
is a
differential of first kind (i.e. holomorphic) on $\Gamma _N$, then $\mu
\omega _j$ is a differential of third kind with simple poles at
$\infty_i$.
The sum of the residues of $\mu \omega _j$ is equal to \begin{equation}
\sum_{i=1}^{N-1} f_j(\infty_i) - (N-1) f_j(\infty_N) = 0 \; .
\label{efj}
\end{equation} Let $\Omega $ be a differential of second kind on $\Gamma
_N$
with a single pole at $\infty_N$. Such is for example the differential
$$
\frac{\mu ^2-\wp(\lambda )}{\frac{\partial f}{\partial \mu } (\lambda
,\mu )}
d\lambda \; .
$$ If moreover $\Omega $ s normalized as
$$
\int_{\tilde{a}_i}\Omega =0
$$ then it is well known that the vector $V$ is co-linear to
$$ (\int_{\tilde{b}_1}\Omega,\int_{\tilde{b}_2}\Omega,...,
\int_{\tilde{b}_N}\Omega)
$$ (see for example [BBEIM 1994]). Equivalently, if we apply the
reciprocity
law to the differentials of second and first kind $\Omega , \omega _i$,
we
get that $V$ is co-linear to
$$ (f_1(\infty_N),f_2(\infty_N),...,f_N(\infty_N)) \; . $$ On the other
hand
$$\tilde{a}_1 = a_1+a_2+...+a_N= \pi^{-1}(a)$$ and hence
$$
\int_{\tilde{a}_1} \omega _i= \sum_{k=1}^N \int_a f_i(\lambda ,\mu _k)
d\lambda
$$ where $(\lambda ,\mu _k)\in \Gamma _N$ are the $N$ pre-images of
$\lambda \in \Gamma_1$. It is clear that $ \sum_{k=1}^N f_i(\lambda ,\mu
_k)$
is a single-valued function on $\Gamma _1$. As $\omega _i$ is a
holomorphic
differential on $\Gamma _N$ and $d\lambda $ is the holomorphic
differential
on $\Gamma _1$, then $\sum_{k=1}^N f_i(\lambda ,\mu _k)$ is a
holomorphic
function on $\Gamma _1$ and hence a constant. As $\omega _i$ is a
normalized
basis of holomorphic forms, then $\int_{\tilde{a}_1}\omega _i=0$ for
$i\geq
2$, and hence $$
\sum_{k=1}^N \int_a f_i(\lambda ,\mu _k) d \lambda =
\sum_{k=1}^N  f_i(\lambda ,\mu _k) \int_a d \lambda
\equiv 0,
\quad (\lambda ,\mu ) \in \Gamma_N, \quad i \geq 2 \;.
$$ Therefore
$$
\sum_{k=1}^N  f_i(\lambda ,\mu _k) \equiv 0, \quad i \geq 2
$$ which combined with (\ref{efj}) implies that
$$ f_i(\infty_N)=0, \quad i\geq 2
$$ and hence the vector $V$ is co-linear to $(1,0,...,0)$. In fact $V$
is
equal to this vector, because $q_i(t)\in \Gamma _1= {\Bbb C}/\{{\Bbb Z}+
\tau
{\Bbb Z}
\}$. Finally, we may always suppose that $U=(0,U_2,...,U_N)$. Indeed the
Calogero system (\ref{calogero}) is invariant under the translation $$
q_i
\rightarrow q_i - V_1 t \; .
$$
$\Box$

{\bf Acknowledgments}. One of authors (A. P.) would like to thank the
Laboratoire Emile Picard at the University of Toulouse III and
Max-Planck-Institut f\"ur Mathematik, Bonn for the hospitality, and B.A.
Dubrovin for the useful remarks.

\end{document}